\documentclass[journal, letterpaper]{IEEEtran}
\IEEEoverridecommandlockouts
\usepackage[dvips]{graphicx}
\usepackage{color}
\usepackage{amssymb,amsmath}
\usepackage{epsfig}
\usepackage{algorithm2e}
\usepackage[compress]{cite}

\begin{document}
\title{Whether and Where to Code in the Wireless Packet Erasure Relay Channel}
\author{\IEEEauthorblockN{Xiaomeng Shi, \textit{Student Member, IEEE}, Muriel M\'{e}dard, \textit{Fellow, IEEE}, and Daniel E. Lucani, \textit{Member, IEEE}}
\thanks{Manuscript received on August 15, 2011, and revised on March 18, 2012.}
\thanks{Xiaomeng Shi (xshi@mit.edu) and Muriel M\'{e}dard (medard@mit.edu) are with the Research Laboratory of Electronics, Massachusetts Institute of Technology, MA, USA. Daniel E. Lucani (dlucani@fe.up.pt) is with the Instituto de Telecomunica\c c\~oes, DEEC Faculdade de Engenharia, Universidade do Porto, Portugal.}
\thanks{The authors acknowledge the financial support of the Interconnect Focus Center, one of the six research centers funded under the Focus Center Research Program, a Semiconductor Research Corporation program. This work is also supported by the NSERC Postgraduate Scholarship (PGS) issued by the Natural Sciences and Engineering Research Council of Canada, and in part by the Funda\c c\~ ao para a Ci\^ encia e a Tecnologia under project PEst-OE/EEI/LA0008/2011.}
}

\maketitle
\begin{abstract}
The throughput benefits of random linear network codes have been studied extensively for wirelined and wireless erasure networks. It is often assumed that all nodes within a network perform coding operations. In energy-constrained systems, however, coding subgraphs should be chosen to control the number of coding nodes while maintaining throughput. In this paper, we explore the strategic use of network coding in the wireless packet erasure relay channel according to both throughput and energy metrics. In the relay channel, a single source communicates to a single sink through the aid of a half-duplex relay. The fluid flow model is used to describe the case where both the source and the relay are coding, and Markov chain models are proposed to describe packet evolution if only the source or only the relay is coding. In addition to transmission energy, we take into account coding and reception energies. We show that coding at the relay alone while operating in a rateless fashion is neither throughput nor energy efficient. Given a set of system parameters, our analysis determines the optimal amount of time the relay should participate in the transmission, and where coding should be performed.
\end{abstract}

\section{Introduction}\label{sec:introduction}
\begin{figure}[t!]
  \centering
  \includegraphics[width=3.2in]{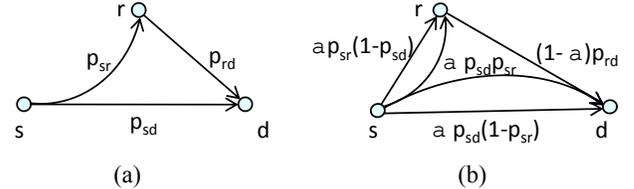}
  \caption{Single relay unicast network, with corresponding flow hypergraph. $p_{sr}$ represents the packet transmission success probabilities between $s$ and $d$.}
  \label{fig:model}
\end{figure}

Network coding, although initially introduced as a theoretical tool in the field of network information theory \cite{ahlswede2000network}, has been made practical by the use of random linear network codes (RLNC) \cite{chou2003practical, ho2006random}, and has been shown to offer throughput, delay, energy and other advantages over classical store-and-forward strategies. To minimize the amount of centralized control, RLNC is often performed at the source as well as all intermediate nodes within a transmission subgraph. Because coding operations and data reception by intermediate nodes can have non-trivial energy costs, resource constrained networks, such as wireless body area networks (WBANs), could potentially benefit from strategies that allow a reduction in the number of coding nodes, while maintaining the benefits of network coding. This paper studies the strategic use of network coding in a three-node wireless packet erasure relay channel, as illustrated by Figure~\ref{fig:model}(a), with an emphasis on whether and where to code when the relay operates in half-duplex mode. We propose Markov chain models to characterize the system performance in terms of throughput and packet delivery energy, thus providing a way to find the optimal fraction of time for which the relay should participate in the transmission. We show, through numerical analysis, that coding at the relay alone while operating in a rateless fashion is neither throughput nor energy efficient, while coding at the source alone has performances close to the case where coding is performed at both nodes. The decision to code is based on packet erasure probabilities, transmission energy, as well as energy spent on reception and coded packet generation.

Although seemingly simple, the analysis of the three-node network can offer insights to more complicated systems with more source or sink nodes. Two such examples are WBAN and advanced LTE cellular networks \cite{3gpp}. In a WBAN, the topology is almost always star-shaped: data are uploaded in a converge-cast sense to a central base station (BS) \cite{latre2011survey,kwak2009study}. Depending on the relative location of a sensor on the human body, it may be useful in terms of energy efficiency to deploy a relay around the shoulder, in direct line of sight with both the front and back of the body. To the best of our knowledge, the throughput and energy tradeoffs in this case have not been studied before. On the other hand, in advanced LTE systems, data are transmitted in both directions, with an emphasis on download in a broadcast sense from a central BS to individual user equipments. Currently relaying is being considered as an improvement tool, for example, for coverage of high data rates, and for temporary network deployment \cite{3gpp}. Here the relay is to wirelessly connect to the radio-access network, and may either function as a smart repeater, or have control of its own cell. For both WBANs and advanced LTE systems, the introduction of network coding and the insertion of a relay may bring energy or throughput gains. As a starting point, we consider the three-node packet erasure relay channel, assuming physical layer designs are readily available on point-to-point links. Further extensions of this setup can involve additional source nodes, as in a WBAN, additional sink nodes, as in an LTE system, and additional relay nodes, a scenario applicable to both examples.

In evaluating system performance, we parametrically model the total energy consumption, taking into account transmission and reception energies as well as processing energy required to generate coded packets. The inclusion of energies for data reception, idle listening, and protocol overheads allows a more comprehensive understanding of the implications of inserting a relay and using network coding. Our goal is to characterize the throughput and energy performances of the system when a relay is added to a point-to-point link and nodes are network coding capable. Our previous work studied the energy advantages of network coding in half-duplex WBANs with a star-topology and showed that when reception energy is taken into account, total energy use could be reduced \cite{shi2011both}. The gains are expected to grow with the number of nodes.

The remaining of the paper is organized as follows. Section~\ref{sec:relatedWork} summarizes some of the previous work related to the wireless relay channel and the application of network coding in such settings. Differently from previous works that focus on joint channel and network coding for optimal throughput analysis, in this paper, we assume a physical layer design is available, and network coding is inserted into the network layer, independently of the source and channel codes employed. Section~\ref{sec:systemModel} details the assumptions made and the system modeled. Section~\ref{sec:analysis} then discusses three separate cases, depending on the coding locations. For each case, we try to characterize the expected completion time and the expected completion energy of transmitting a given number of packets from the source to the destination. Through Markov chain analysis, we provide a framework for evaluating bounds on the system performance when coding is conducted at the relay only, and for determining the system performance when coding is conducted at the source only. Section~\ref{sec:simulations} then establishes the key results through numerical evaluations, showing that, for a wide range of parameters, coding at the relay only is not throughput or energy efficient. Section~\ref{sec:conclusion} concludes the paper with discussions on future work.

\section{Related Work}\label{sec:relatedWork}
A relay channel models the problem where two nodes communicate through the help of one or more relays. This setup is common in multihop wireless networks such as sensor networks, where transmission power is limited, or in decentralized ad hoc networks, where nodes can only communicate with their immediate neighbors. Relays can overhear transmissions to the destination, owing to the broadcast advantage of the wireless medium. Recently there has been a renewed interest in the classical relay channel \cite{van1968transmission, cover1979capacity}, motivated by the potential to achieve cooperative diversity, and thus better capacity bounds \cite{laneman2004cooperative, kramer2005cooperative,lai2006three, dana2006capacity}. Schemes such as amplify-and-forward, decode-and-forward, and compress-and-forward have been proposed and studied extensively in terms of capacity, outage, energy efficiency, and optimal power allocation schemes \cite{yao2005energy, zhao2007improving}. Much of the analysis has focused on the fundamental performance limits at the physical layer and on the transmission of a single data packet. The introduction of network coding into the relay channel has also focused mostly on joint channel-network code design, with or without limited processing complexity constraints \cite{hausl2006iterative, xiao2007network, zhang2009channel, yang2007network, tuninetti2005throughput, tuninetti2004processing}. In larger networks, however, network coding typically resides in higher layers of the protocol stack, independently of physical layer implementations.

In this paper, we assume network coding takes place at the protocol layer, independently of source and channel codes employed. Such an assumption on the separation of channel and network codes may not necessarily be capacity achieving, but allows the introduction of network coding into existing systems.

The use of RLNC in wireless erasure networks under packetized operations is first studied by Lun et al. \cite{lun2008coding}, and extended to a scheduling framework by Traskov et al. \cite{traskovThesis}. Other schemes that employ network coding in a relay setup includes the MORE protocol \cite{chachulski2007trading}, which performs RLNC at the source only to reduce the amount of coordination required by multiple relay nodes, and the COPE protocol, which employs RLNC at the relay only in a 2-way relay channel to improve reliability, taking advantage of opportunistic listening and coding \cite{katti2006xors}. Fan et al. also proposed a network coding based cooperative multicast scheme to show that significant throughput gains can be achieved when network coding is performed at the relay only \cite{fan2009reliable}; one assumption in this work is that feedback is available from both the destination and the relay to the source after each packet reception. In practical systems, feedback can be costly in terms of both throughput and energy, depending on the underlying hardware architecture \cite{shi2011both}.

In this paper, we explore rateless transmissions, where the acknowledgement for successful reception is sent only once by the destination when the transmission of all available data is completed. As described in the introduction, we also take into account the energy spent on reception and packet processing in addition to the energy required to transmit them. Furthermore, we assume that a sufficiently large field is used for network coding operations, such that transmissions of non-innovative packets from the source can be neglected. In terms of energy use, we make the simple assumption that coding energy stays constant as field size increases, and show that, the decision to code depends on the dominating energy term (transmission, reception, or code generation). The tradeoff between energy budget for the transmission of linearly dependent packets when field size is small and the energy budget for code generation when field size is large is discussed in \cite{angelopoulos2011energy}.

\section{System Model}\label{sec:systemModel}
We represent the data flow through the relay channel using a hypergraph, as shown in Figure~\ref{fig:model}(a). A hypergraph is a generalization of a graph: a broadcast link is represented by a hyperarc between a single start node and a set of end nodes, and a multiple access link is represented by a hyperarc between a set of start nodes and a single end node \cite{lun2008coding}. A wireless relay channel consists of a source node $s$, a relay node $r$, and a sink node $d$. Source $s$ has $n$ packets of the same length to transmit to $d$. It broadcasts to both $r$ and $d$, while the relay $r$ assists the transmission by either forwarding the original packet, or computing linear combinations of received packets before forwarding the ensuing mixtures.

We assume transmissions occur in a rateless fashion, with minimal feedback: $s$ and $r$ take turns to transmit, until $d$ acknowledges that it has received enough degrees of freedom (dof) to recover the original $n$ data packets. Here we use dofs to represent linearly independent packets. Such rateless operations are often desirable in systems where feedback can be costly in terms of energy or delay.

Our model considers packetized operations, independently of the physical layer implementation of the system. As such, erroneous packets are dropped, and channel losses are measured by a time-averaged erasure rate. This separation of channel and network coding follows from the assumption that physical layer designs are already available for the underlying point-to-point link, with a relay being inserted for performance improvements. The transmission success rates are assumed to be $p_{sr}$ between $s$ and $r$, $p_{rd}$ between $r$ and $d$, and $p_{sd}$ between $s$ and $d$. Nodes operate in half-duplex mode, where a node cannot transmit and receive at the same time. To avoid interferences and collisions in a contention based scheme, we consider a time-division framework, where $s$ and $r$ share the use of the wireless medium. A genie scheduler allocates the wireless medium to the source $\alpha$ fraction of the total time, $0\leq \alpha \leq 1$, and allocates the wireless medium to the relay the remaining $1-\alpha$ fraction of time. One possible implementation of such a genie-aided scheduler is to share the same randomness at $s$ and $r$. Figure~\ref{fig:model}(b) illustrates the maximum flow on each possible link in this network model, computed directly from the transmission success rates and the time-sharing constant $\alpha$.

In terms of memory, let both $s$ and $d$ contain $n$ units, but assume $r$ contains $x$ units only, where $1\leq x \leq n$. $r$ uses its memory as a queue: arriving packets are stored; if $r$ is already full, newly arrived packets are discarded. If $r$ does not perform coding, it sends to $d$ a packet from its memory directly and drops this packet from the queue; if $r$ performs RLNC before forwarding, it sends to $d$ a linear combination of stored packets, where each is weighted by a random number chosen uniformly from a finite field $\mathbb{F}_q$. In this paper, $\mathbb{F}_q$ is assumed to be sufficiently large, such that linear combinations thus generated are linearly independent from each other with high probability. Reference \cite{angelopoulos2011energy} discusses the tradeoff between field size and energy use in more details.

To evaluate the amount of energy spent to deliver successfully a packet from $s$ to $d$, we define four different energy terms. $E_{tr}$ represents the transmission energy per packet, where transmission occurs at either $s$ or $r$. $E_{rx}$ represents the reception energy per packet at $r$. The relay therefore pays for being on and listening to the broadcast from the source. $E_{nc}$ represents energy for generating a coded packet; it should be a function of $n$, since the complexity of network coding operations depends on field size and generation size, which is the number of packets coded together. Nonetheless, in this paper, $E_{nc}$ is assumed to be constant, representing a maximum allowable value. Lastly, $E_{ack}$ represents the amount of energy spent by $s$ to listen to the final acknowledgement from $d$. Note that all energy terms are defined relative to $s$ or $r$. It is assumed that the destination $d$ represents a base station without power or energy constraints.

\section{Network coding in the Wireless Relay Channel}\label{sec:analysis}
In this section, three different cases are examined: RLNC at both $r$ and $s$, RLNC at $r$ alone, and RLNC at $s$ alone. For the first case, a fluid flow model is used to analyze the achievable rate, packet delivery energy, and the ratio of these two metrics. For the latter cases, we propose Markov chain models to characterize the expected completion time and the expected completion energy of transmitting $n$ packets from $s$ to $d$. We also offer a brief discussion on the use of systematic codes, which will be studied in future works.

\subsection{Coding at Both the Source $s$ and the Relay $r$}\label{subsec:rs}

When RLNC is performed at both $s$ and $r$, we can use the fluid flow model by Lun et al. \cite{lun2008coding, traskov2008scheduling} to study the rateless transmission of network coded packets through the relay channel. A packet is considered innovative when it carries a new dof to a node. In the relay channel, $s$ injects innovative packets into the hyperarc $(s, \{r,d\})$ in $\alpha$ fraction of the total transmission time, while $r$ injects innovative packets into the arc $(r,d)$ in $1-\alpha$ fraction of the total transmission time. Packet transmissions form innovative flows in this setup because both $s$ and $r$ perform RLNC over a large number of packets. Each mixture is an additional dof relative to $d$. The amount of innovative flow is limited by the packet erasure probabilities. Assuming flow conservation at $r$, the maximum achievable rate $R$ from $s$ to $d$ can be derived by solving the following mathematical programming problem analytically using Fourier-Motzkin elimination \cite{traskovThesis}.
\begin{align*}
 \min \quad & c(R,\alpha) \\
 \mbox{s.t.} \quad
 & R \leq \alpha (p_{sr} + p_{sd} - p_{sr}p_{sd}) \\
 & R \leq \alpha p_{sd} + (1-\alpha)p_{rd}\\
 & 0 < \alpha \leq 1\,.
\end{align*}

\subsubsection{Packet-Level Capacity Bound}\label{subsucsec:traskov} $c(R,\alpha)=1/R$.
This case is equivalent to maximizing $R$, making the optimization linear. A closed-form solution can be found:
\begin{itemize}
 \item Case 1: $p_{sd} \leq p_{rd}$, then
  \begin{align*}
    R^*        & = \frac{p_{rd}(p_{sr} + p_{sd} - p_{sd}p_{sr})}{p_{rd}+p_{sr}(1-p_{sd})}\,, \\[4pt]
    \alpha ^*  & = \frac{p_{rd}}{p_{rd}+p_{sr}(1-p_{sd})}\,.
  \end{align*}
 \item Case 2: $p_{sd} > p_{rd}$, then the relay is not used, and
  \begin{align}
    R^* &       = p_{sd}\,, \quad \alpha^*=1\,. \label{eq:alpha}
  \end{align}
\end{itemize}

\subsubsection{Packet-Delivery Energy}\label{subsucsec:pktDevliveryE} $c(R,\alpha) = [E_{tx}+E_{nc}+\alpha (1 - \mathbf{I}_{\alpha=1})E_{rx}]/R$. We define the packet delivery energy as the energy consumed per successfully transmitted packet and state it explicitly in terms of packet transmission, reception, and coding energies. $\mathbf{I}_{\alpha=1}$ represents an indicator function that equals to 1 if $\alpha=1$, and 0 otherwise. In other words, if $\alpha=1$, the relay is not used thus should not consume any energy. Because the scheme considered is rateless in the limit of infinitely large payloads, no energy is spent on listening to the acknowledgement at $s$. Also observe that if $E_{rx} =0$, this problem reduces to case $1)$, and the optimal $\alpha^*$ which achieves the highest throughput also leads to a minimal packet delivery energy. If $E_{rx}$ is non-zero, we can rewrite the optimization in terms of $\alpha$ alone and solve numerically.

\subsubsection{Packet-Delivery Energy per Throughput Rate}\label{subsubsec:pktDeliveryEperThrpt}
$c(R,\alpha) = [E_{tx}+E_{nc}+\alpha (1 - \mathbf{I}_{\alpha=1})E_{rx}])/R^2$. The ratio of the two metrics above is the expected packet delivery energy per throughput rate, which evaluates changes in both energy and throughput.
\subsection{RLNC at the Relay $r$ Only}\label{subsec:r}
First assume that $s$ is limited to sending the original packets only, while $r$ performs RLNC over all packets it has received and stored in memory. Since transmission is rateless with minimal feedback, $s$ is unaware of the knowledge at $r$ or $d$; such knowledge refers to specific dofs, representing a subspace spanned by received packets. We assume that when allowed to transmit, $s$ chooses one packet uniformly at random from the $n$ available uncoded packets.

One remark here is that more exhaustive or frequent feedbacks would make retransmissions from $s$ more likely to be innovative. For example, if per-packet acknowledgement is available, $s$ can selectively repeat those not successfully received by $r$ or $d$. On the other hand, if $d$ can acknowledge the exact number of dofs received, $r$ can adjust the number of coded packets it sends to maximize throughput to $d$ while minimizing its own energy use. We consider transmission with minimal feedback, assuming that the cost of feedback is high.

We describe the states of the system by a three-tuple $(m,k,l)$: $m$ is the number of unique dofs at $d$, $k$ is the number of dofs shared by $d$ and $r$, and $l$ is the number of dofs at $r$ only. Since $s$ does not code and $r$ stores only uncoded packets, $m$ and $l$ represent the numbers of distinct packets at $r$ and $d$ respectively; e.g., if $n=3$, $r$ has successfully received packets $1$ and $2$, while $d$ has received packets $2$ and $3$, then $(m,k,l)=(1,1,1)$. Once a packet at $r$ has been mixed into a coded packet, which in turn is received at $d$, it becomes part of the shared dof between $r$ and $d$; e.g., if $r$ has received packets $1$ and $2$, while $d$ has packet $3$ and a mixture of packets $1$ and $2$, then $(m,k,l) = (1,1,1)$, since the mixture represents a shared dof between $r$ and $d$, while $r$ contains one dof that is innovative to $d$. With such state definitions, any three-tuple satisfying $m+k+l\leq n$ is a valid state. Transmission initiates in state $(0,0,0)$, and terminates in states $\{(m^*,k^*,l^*)|\,m^*+k^*=n\}$.

For transmissions to be free of collisions, recall from the system model that we assume there is a genie-aided scheduler, such that $s$ and $r$ do not access the wireless medium at the same time. In each time slot, $s$ transmits with probability $\alpha$, while $r$ transmits with probability $1-\alpha$. By randomizing the transmitter at each time slot, the state transition process becomes memoryless, and the numbers of dof at each node can be tracked through a Markov chain. The memoryless property holds because the probability of the next transmitted packet being innovative relative to both or either of $r$ and $d$ can be expressed in terms of $\alpha$ and the current state $(m,k,l)$, independently of past state evolutions. With probability $\alpha$, $s$ chooses one packet uniformly at random from its $n$ uncoded packets, and with probability $(n-m-k-l)/n$, this packet is innovative to both $r$ and $d$. Similarly, with probability $1-\alpha$, $r$ computes a linear combination of the content of its memory, and sends the mixture to $d$. Here we assume $r$ has enough memory to store all $n$ distinct packets.

An alternative to the genie-aided randomized transmissions is a collision-free, deterministic schedule, where $s$ and $r$ take turns to transmit for a fixed amount of time, determined by $n$, $\alpha$, and the channel conditions. Without feedback, the average system throughput should be the same as the randomized case. However, with this deterministic schedule, since packets are uncoded at $s$, counting the numbers of dof at $r$ and $d$ requires knowledge of the exact packets present. Even with a small $n$, it is hard to track the evolution of packets in the system.
\begin{figure}[t!]
  \centering
  \includegraphics[width=2.7in]{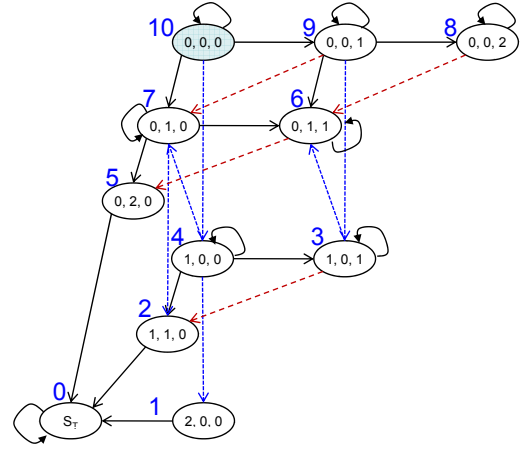}
  \caption{Markov chain model, with added terminating state $S_T$. The number of packets to send at $s$ is $n = 2$. Any three-tuple $(m,k,l)$ satisfying $m+k+l\leq n$ is a valid state.}
  \label{fig:mc_n2}
\end{figure}

Figure~\ref{fig:mc_n2} gives a sample Markov chain when $n=2$. We have drawn the Markov chain as a tetrahedron, with the starting state on top, and the terminating states on the left vertical edge. State transitions occur after the transmission of a single packet, either from $s$ or from $r$. In the case where $r$ has not received any packet successfully but is chosen to transmit, we assume the slot is wasted. At state $(m,k,l)$, the transition probabilities can be computed by considering all outcomes of the transmission, assuming independent packet losses.

When $m+k < n$, $s$ broadcasts with probability $\alpha$, while $r$ transmits with probability $1-\alpha$. Let the indicator function $\mathbf{I}_{l>0}$ be 1 when $l\neq0$, and $0$ otherwise, then
\begin{align}
    P_{\{m,k,l\}\rightarrow\{m+1,k,l\}}   &= \frac{n-m-k-l}{n}p_{sd}(1-p_{sr})\alpha \label{eq:r1}\\
    P_{\{m,k,l\}\rightarrow\{m,k+1,l\}}   &= \frac{n-m-k-l}{n}p_{sd}p_{sr}\alpha \label{eq:r2}\\
    P_{\{m,k,l\}\rightarrow\{m,k,l+1\}}   &= \frac{n-m-k-l}{n}p_{sr}(1-p_{sd})\alpha \label{eq:r3}\\
    P_{\{m,k,l\}\rightarrow\{m-1,k+1,l\}} &= \frac{m}{n}p_{sr}\alpha\label{eq:r4} \\
    P_{\{m,k,l\}\rightarrow\{m,k+1,l-1\}} &= \frac{l}{n}p_{sd}\alpha+\mathbf{I}_{l>0}p_{rd}(1-\alpha)\label{eq:r5}
\end{align}
\begin{align}
    P_{\{m,k,l\}\rightarrow\{m,k,l\}}   &=  \left[\frac{m}{n}(1-p_{sr})+\frac{l}{n}(1-p_{sd})
     + \frac{k}{n} \right. \notag \\ & \left. + \frac{n-m-k-l}{n}(1-p_{sd})(1-p_{sr})\right]\alpha \notag \\
     & \hspace{.1in} + (1-\mathbf{I}_{l>0}p_{rd})(1-\alpha) \label{eq:r6}
\end{align}

When $s$ broadcasts, this transmission may succeed in any of the three hyperarcs originating from $s$, as shown in Figure~\ref{fig:model}(b). Depending on which packets $r$ and $d$ already have, a successful transmission may or may not lead to a transition to a different state. For example, if the transmitted packet has already been received by both nodes, a state transition does not occur regardless of whether the transmission is successful. In particular, the following state transitions are possible.

\begin{enumerate}
    \item If $d$ receives one more dof, while $r$ does not, $P_{\{m,k,l\}\rightarrow\{m+1,k,l\}} = \frac{n-m-k-l}{n}p_{sd}(1-p_{sr})$, this term is included in Eq.~\eqref{eq:r1};
    \item if both $d$ and $r$ receive one more dof, $P_{\{m,k,l\}\rightarrow\{m,k+1,l\}} = \frac{n-m-k-l}{n}p_{sd}p_{sr}$, i.e., Eq.~\eqref{eq:r2};
    \item if $r$ receives one more dof, while $d$ does not, $P_{\{m,k,l\}\rightarrow\{m,k,l+1\}} = \frac{n-m-k-l}{n}p_{sr}(1-p_{sd})$, i.e., Eq.~\eqref{eq:r3};
    \item if the broadcasted packet has been previously received by $d$, and is now received at $r$, $P_{\{m,k,l\}\rightarrow\{m-1,k+1,l\}} = \frac{m}{n}p_{sr}$; note that if $m=0$, this transition probability is 0, i.e., Eq.~\eqref{eq:r4};
    \item if the broadcasted packet has been received by $r$ previously, and is now received at $d$, then $P_{\{m,k,l\}\rightarrow\{m,k+1,l-1\}} = \frac{l}{n}p_{sd}$; if $l=0$, this transition probability is 0, i.e., Eq.~\eqref{eq:r5};
    \item a self transition occurs if the dof being sent has previously been received by $d$ but not $r$, if it has previously been received by $r$ but not $d$, if it is already shared, or if it has not been shared previously, yet it is not received successfully by $r$ nor $d$ during this transmission; correspondingly,
    $P_{\{m,k,l\}\rightarrow\{m,k,l\}}$ is the sum of four terms: $\frac{m}{n}(1-p_{sr})$, $\frac{l}{n}(1-p_{sd})$, $\frac{k}{n}$, and $\frac{n-m-k-l}{n}(1-p_{sd})(1-p_{sr})$, i.e., Eq.~\eqref{eq:r6}.
\end{enumerate}

$r$ transmits coded packets with probability  $1-\alpha$. Observe that, when $r$ has received only a small number of packets, a mixture it generates may not be innovative with respect to $d$, because the dof could have been received by $d$ itself already. For example, if $n$ is equal to 3, and $d$ has already received packets 2 and 3, then a coded packet from $r$ containing the weighted sum of packets 2 and 3 is not innovative even if it is successfully received at $d$, but a coded packet containing the weighted sum of packets 1 and 2 is innovative.

Again, explicitly tracking the contents of coded packets is a difficult task. Instead, we assume that all packets transmitted from $r$ to $d$ are innovative. It is important to note that the computed expected completion time under this assumption is a \emph{lower bound} on the actual expected completion time, and the corresponding throughput is an \emph{upper bound} on the actual system throughput. When discussing numerical results in Section~\ref{sec:simulations}, we shall show that even this upper bound on throughput is \emph{not} efficient compared with schemes where coding is performed at $s$, or at both $s$ and $r$. Also observe that if $n=1$, an uncoded packet is always transmitted; the innovative packet assumption is then always true, with the computed expected completion time being exact. With such assumptions, the following state transitions can occur

\begin{enumerate}
    \item if $r$ has no unique dof to share, $l=0$, $P_{\{m,k,0\}\rightarrow\{m,k,0\}} = 1$, i.e., Eq.~\eqref{eq:r6};
    \item if $r$ has a unique dof to share, $l>0$, and $d$ receives successfully,
            $P_{\{m,k,l\}\rightarrow\{m,k+1,l-1\}} = p_{rd}$, i.e., Eq.~\eqref{eq:r5};
    \item if $r$ has a unique dof to share, $l>0$, but $d$ does not receive successfully, $P_{\{m,k,l\}\rightarrow\{m,k,l\}} = 1-p_{rd}$, i.e., Eq.~\eqref{eq:r6}.
\end{enumerate}

The transmission process terminates when $m+k=n$, and $P_{\{m,n-m,0\}\rightarrow\{m,n-m,0\}}=1$. In this Markov chain, all states are transient except the absorbing states. When $n>1$, multiple recurrent classes exist. Since there is a single starting state, there exists a unique steady state distribution. To simplify the computation of the absorbing time, we append a virtual terminating state $S_T$, such that $P_{\{m,n-m,0\}\rightarrow\{S_T\}}=1$, and $P_{\{S_T\}\rightarrow\{S_T\}}=1$. To compute explicitly the transition state matrix, we can index the states linearly starting from $S_T$ to $(0,0,0)$. Let $S_T$ be state 0 under this counting notation. In Figure~\ref{fig:mc_n2}, we give one possible set of linear indices, starting from the bottom to the top of the tetrahedron.

Let $T_i$ be the expected first passage time to state 0, i.e., the expected number of steps to reach state 0, starting in state $i$. Also let $T$ be the system expected transmission completion time. Hence $T=T_{10}$ in this example. Since there are no cycles in this Markov chain and the expectation operation is linear, $T_i$ can be solved recursively using the equation
\begin{align*}
T_{i} &= \frac{1}{1-P_{ii}} \left\{ 1 + \sum_{j\neq i} P_{ij} T_{j} \right\}, \quad T_{0} = 0\,, i\neq 0.
\end{align*}
Alternatively, if we express the expected first passage times to state $0$ in vector form $\bar{T} = \left[
                                                                           \begin{array}{cccc}
                                                                             T_1 & T_2 & \ldots & T_{10} \\
                                                                           \end{array}
                                                                         \right]^T
$, then $\bar{T} = \mathbf{1}+P_{\backslash 0}\bar{T}$, where $\mathbf{1}$ is the vector of ones, and  $P_{\backslash 0}$ is the submatrix of $P$ with the first row and first column removed. Solving this linear system of equations gives
\begin{align}
 \bar{T} &= (\mathbf{I} - P_{\backslash 0})^{-1} \mathbf{1}\,.\label{eq:T}
\end{align}
Since the original Markov chain does not contain cycles other than loops, the states have a topological order; $P$ is a lower-triangular matrix. In addition, since self-transition probabilities are non-zero for all states except those next to $S_T$, $I-P_{\backslash 0}$ is strictly lower-triangular, thus invertible in the real field.

To determine the expected amount of energy consumed by $s$ and $r$, let $E_i$ be the expected energy to be computed, staring from state $i$, and $E$ be the expected transmission completion energy, i.e., $E=E_{10}+E_{ack}$. Here $E_{ack}$ is included since a single acknowledgement is sent by $r$ at the end to signal the end of transmission. A similar argument holds as in the derivation of Eq.~\eqref{eq:T}, and a system of linear equation can be solved to find $\bar{E} = \left[
                                                                           \begin{array}{cccc}
                                                                             E_1 & E_2 & \ldots & E_{10} \\
                                                                           \end{array}
                                                                         \right]^T =(\mathbf{I} - P_{\backslash 0})^{-1} (\mathbf{1}E_{use}) = \bar{T}E_{use}
$, where $E_{use}= E_{tx} + \alpha(1-\mathbf{I}_{\alpha=1})E_{rx}+(1-\alpha)E_{nc}$.

One last observation is that it is possible to constrain the amount of memory at $r$ to less than $n$, and have $r$ function as an accumulator, such that whenever an innovative packet is received, it is multiplied by new random coefficients and added to each of the memory units. When given a transmission opportunity, $r$ uniformly randomly chooses one mixture from its memory. The achievable rate region of the limited memory case should be outer-bounded by the full memory case. We shall show in Section~\ref{sec:simulations} that even with full memory, coding at $r$ alone is not efficient.


\subsection{RLNC at the Source Only}\label{subsec:s}

When RLNC is performed at $s$ only, the analysis of the system performance is similar to the previous case, where RLNC is performed at $r$ only. First, every mixture sent by $s$ is innovative with respect to $r$ and $d$ under the infinite field size assumption. Let state $(m,k,l)$ represent $m$ unique mixtures at $d$, $k$ mixtures shared by $r$ and $d$, and $l$ unique mixtures at $r$. Assume $r$ acts as a queue with $x$ finite units of memory, where $1\leq x\leq n$. Here $r$ is allowed to have fewer than $n$ units of memory, since it does not need to store distinct packets for explicit coding operations. $r$ receives linear mixtures directly from $s$, functions as a queue, and drops any mixtures received after it is full. Any mixture transmitted from $r$ is also dropped from the queue. A state $(m,k,l)$ is valid as long as $m+k\leq n$ and $k+l\leq x$. Again, $s$ transmits $\alpha$ fraction of the time. We assume that $P_{\{m,k,l\}\rightarrow\{m',k',l'\}}$ is non-zero only if both $(m,k,l)$ and $(m',k',l')$ are valid states. Let the indicator function $\mathbf{I}_{f(\cdot)}$ be 1 if the logic function $f(\cdot)$ is true, and 0 otherwise. State transitions occur after the transmission of a single packet, either from $s$ or from $r$. The state transition probabilities are as follows,
\begin{align}
    P_{\{m,k,l\}\rightarrow\{m+1,k,l\}}   &= \alpha p_{sd}(1-p_{sr}) + \alpha p_{sd}p_{sr} \mathbf{I}_{k+l=x} \label{eq:s1}\\
    P_{\{m,k,l\}\rightarrow\{m,k+1,l\}}   &= \alpha p_{sd}p_{sr} \mathbf{I}_{k+l<x} \label{eq:s2}\\
    P_{\{m,k,l\}\rightarrow\{m,k,l+1\}}   &= \alpha p_{sr}(1-p_{sd}) \mathbf{I}_{k+l<x} \label{eq:s3}\\
    P_{\{m,k,l\}\rightarrow\{m,k,l\}}     &= \alpha p_{sr}(1-p_{sd})\mathbf{I}_{k+l=x} \notag \\ & \hspace{-0.5in}+ \alpha (1-p_{sr})(1-p_{sd})+ (1-\alpha)\mathbf{I}_{k+l=0} \label{eq:s4}\\
    P_{\{m,k,l\}\rightarrow\{m+1,k-1,l\}} &= (1-\alpha) \frac{k}{l+k} \mathbf{I}_{k>0} \label{eq:s5}\\
    P_{\{m,k,l\}\rightarrow\{m+1,k,l-1\}} &= (1-\alpha) \frac{l}{l+k} p_{rd}\mathbf{I}_{l>0} \label{eq:s6}\\
    P_{\{m,k,l\}\rightarrow\{m,k,l-1\}}   &= (1-\alpha) \frac{l}{l+k} (1-p_{rd}) \mathbf{I}_{l>0} \label{eq:s7}
\end{align}
In the case where $r$ has not received any packet successfully but is chosen to transmit, the slot is assumed to be wasted. Assuming independent packet losses, the state transition probabilities are computed as described below.

First, when $m+k < n$, $s$ broadcasts with probability $\alpha$, and the following can occur.

\begin{enumerate}
    \item If $d$ receives the transmitted mixture, but $r$ does not, $P_{\{m,k,l\}\rightarrow\{m+1,k,l\}} = p_{sd}(1-p_{sr})$, i.e., Eq.~\eqref{eq:s1};
    \item if both $d$ and $r$ receive the transmitted mixture, and $k+l<x$, $P_{\{m,k,l\}\rightarrow\{m,k+1,l\}} =  p_{sd}p_{sr}$, i.e., Eq.~\eqref{eq:s2};
    \item if both $d$ and $r$ receive the transmitted mixture, and $k+l=x$, $P_{\{m,k,l\}\rightarrow\{m+1,k,l\}} =  p_{sd}p_{sr}$, i.e., Eq.~\eqref{eq:s1};
    \item if $r$ receives the transmitted mixture, but $d$ does not, and $k+l<x$, the mixture is stored in memory, $P_{\{m,k,l\}\rightarrow\{m,k,l+1\}} = p_{sr}(1-p_{sd})$, i.e., Eq.~\eqref{eq:s3};
    \item if $r$ receives the transmitted mixture, but $d$ does not, and $k+l=x$, the mixture is dropped, $P_{\{m,k,l\}\rightarrow\{m,k,l\}} = p_{sr}(1-p_{sd})$, i.e., Eq.~\eqref{eq:s4};
    \item if neither $r$ nor $d$ receives the packet, $P_{\{m,k,l\}\rightarrow\{m,k,l\}} = (1-p_{sr})(1-p_{sd})$, i.e., Eq.~\eqref{eq:s4}.
\end{enumerate}

\noindent $r$ transmits coded packets with probability $1-\alpha$, and the following state transitions can occur.
\begin{enumerate}
    \item If $r$ has no unique mixture to share, $l=0$, and $k=0$, $P_{\{m,k,l\}\rightarrow\{m,k,l\}}= 1$, i.e., Eq.~\eqref{eq:s4};
    \item if $r$ has no unique mixture to share, $l=0$, and $k>0$, $P_{\{m,k,l\}\rightarrow\{m+1,k-1,l\}} = 1 $, , i.e., Eq.~\eqref{eq:s5}; observe that since a packet is dropped from $r$'s memory after being sent, $k$ decrements by 1 since the dof is no longer shared, while $m$ increments by 1 since this dof becomes unique to $d$;
    \item if $r$ has a unique mixture to share, $l>0$, and
        \begin{itemize}
        \item a unique mixture is sent, $d$ receives successfully,
            $P_{\{m,k,l\}\rightarrow\{m+1,k,l-1\}} = \frac{l}{l+k}p_{rd}$, i.e., Eq.~\eqref{eq:s6};
        \item a unique mixture is sent, transmission is unsuccessful,
            $P_{\{m,k,l\}\rightarrow\{m,k,l-1\}} = \frac{l}{l+k}(1-p_{rd})$, i.e., Eq.~\eqref{eq:s7};
        \item $k>0$, a shared mixture is sent, $d$ receives successfully,
            $P_{\{m,k,l\}\rightarrow\{m+1,k-1,l\}} = \frac{k}{l+k}p_{rd}$, i.e., Eq.~\eqref{eq:s5};
        \item $k>0$, a shared mixture is sent, transmission is unsuccessful,
            $P_{\{m,k,l\}\rightarrow\{m+1,k-1,l\}} = \frac{k}{l+k}(1-p_{rd})$, i.e., Eq.~\eqref{eq:s5}.
        \end{itemize}
\end{enumerate}

\begin{figure}[t!]
  \centering
  \includegraphics[width=2.7in]{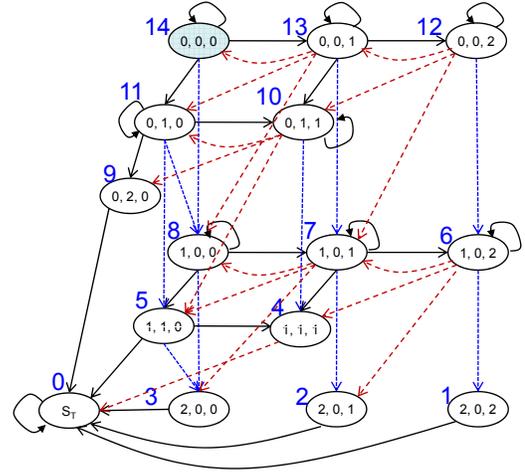}
  \caption{Markov chain model, coding at the source $s$ only,  with added terminating state $S_T$. The number of packets to send at $s$ is $n = 2$; the amount of memory at the relay $r$ is $x=2$.}
  \label{fig:mc_n2_sOnly}
\end{figure}

Transmission terminates when $m+k=n$, and $P_{\{m,n-m,0\}\rightarrow\{m,n-m,0\}}=1$. Again, a virtual terminating state $S_T$ can be appended, such that $P_{\{m,n-m,l\}\rightarrow\{S_T\}}=1$, and $P_{\{S_T\}\rightarrow\{S_T\}}=1$. With this addition, the Markov chain has one recurrent state only. Figure~\ref{fig:mc_n2_sOnly} gives a sample Markov chain when $n=2$. The states can be indexed linearly starting from $S_T$ as state 0, to $(0,0,0)$ as the last state, which corresponds to state number 14 in this example.

Our goal is to find the value of $\alpha$ that minimizes either the expected completion time $T=T_{14}$, or the expected completion energy $E=E_{14}$. Unlike the coding at $r$ only case, the Markov chain now contains cyclic paths in addition to loops. The expected first passage time starting from different states is $\bar{T} = (\mathbf{I} - P_{\backslash 0})^{-1} \mathbf{1}$. Here, the invertibility of $\mathbf{I} - P_{\backslash 0}$ is guaranteed because  $P_{\backslash 0}$ has entries less than $1$ on the main diagonal, and $\mathbf{I} - P_{\backslash 0}$ is a lower Hessenberg matrix with non-zero entries on the main diagonal \cite{horn1990matrix}. Once the value of $\alpha$ that minimizes $T$ is found, we can compute the associated $E$, where $E=TE_{use}+E_{ack}$. The only difference is the value of $E_{use}$, since coding is now performed at the source $s$: $E_{use}= E_{tx} + \alpha(1-\mathbf{I}_{\alpha=1})E_{rx}+(1-\alpha)E_{nc}$.

\subsection{Use of Systematic Codes}\label{subsec:systematic}

Systematic network codes are an attractive alternative to non-systematic random linear network codes, since they often reduce computation complexity and energy use, while maintaining the innovation of independent flows \cite{lucani2010systematic}. With a systematic code at $s$ only, $s$ can first broadcast the uncoded packets one by one in order, then compute random linear mixtures for all remaining packets transmitted from $s$. $r$ performs the store-and-forward function always. In a sufficiently large finite field, since every packet sent by $s$ is innovative with respect to $r$ and $d$, if we view each uncoded packet as an innovative mixture, the state evolution under this setup is the same as the case where full RLNC is performed at $s$ only. If systematic coding is performed at $s$ and RLNC is performed at $r$, the system gives the same performance as the case where RLNC is performed at both nodes. Another possibility is to perform systematic coding at both $s$ and $r$. $s$ first broadcasts uncoded packets one by one in order. It then computes a random linear mixture of all $n$ packets whenever a transmission opportunity becomes available. $r$ acts as a size $n$ queue. When the relay has the opportunity to transmit, it examines the next packet in the queue. If this packet is uncoded, $r$ transmits the uncoded packet directly. If this packet is coded, $r$ linearly combines all data it has in memory before sending out the mixture to $d$. The system performance under this setup should be upper-bounded by the full coding case, and lower-bounded by the coding at $s$ only case. The analysis of this additional systematic phase is non-trivial, so we leave its description and discussion to a later time.

\section{Numerical Results}\label{sec:simulations}

This section compares the performance of the three schemes discussed in Section~\ref{sec:analysis} under different channel conditions. We first consider the coding at $r$ only and coding at $s$ only cases and examine the expected transmission completion times per data packet. We then compare the three cases in terms of achievable throughput, computed as the inverse of average completion time, and packet delivery energy.

\subsection{RLNC at the Relay $r$ Only}

\begin{figure}[t!]
  \centering
  \includegraphics[width=3.2in]{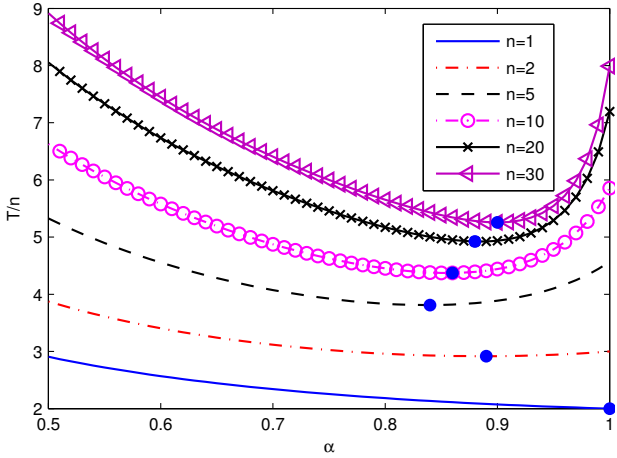}
  \caption{Coding at the relay $r$ only, expected completion time per packet $T/n$  vs. $\alpha$, as $n$ changes in value; $p_{sd} = 0.5$, $p_{sr}=0.8$, $p_{rd} =0.8$. The optimal $T^*/n$ is labeled with a large dot on each curve.}
  \label{fig:rOnly_n}
\end{figure}

Figure~\ref{fig:rOnly_n} plots the expected completion time per transmitted data packet as a function of $\alpha$ for different values of $n$, when $p_{sd}=0.5$, $p_{sr}=0.8$, and $p_{rd}=0.8$. Recall that $n$ represents the number of data packets to be transmitted by the source to the destination. The optimal $\alpha^*$ value that achieves the lowest $T^*/n$ is indicated by a large dot on each curve.

In this figure, when $\alpha = 1$, $r$ listens but does not transmit. If $n=1$, the expected number of transmissions per data packet is 2. This is the solution to the ARQ scheme when $p_{sd} = 0.5$, where each packet is retransmitted until successfully received at $d$. When $n=2$, the expected number of transmissions per data packet is 3. Observe that, since $r$ is unused and $s$ does not code, $s$ simply retransmits one of the two uncoded data packets each round, until both are received at $d$. This scenario is similar to the coupon collector's problem with 2 coupons, except packet erasures need to be taken into account. When 2 coupons are to be collected, the expected number of trials until success is $2\times(1+\frac12) =3$. If divided by $p_{sd}$ and normalized by the number of packets, this solution leads to the value of 3, the value on the curve $n=2$, at $\alpha=1$, in Figure~\ref{fig:rOnly_n}.

Another observation from this figure is that, as $n$ increases, the expected completion time $T/n$ increases as well. This increase comes from transmissions by $s$. Since $s$ randomly chooses one from $n$ packets to transmit, a packet to be transmitted would have been received by $r$ or $d$ already with non-zero probability. This effect is especially significant towards the end of the transmission, when $d$ has collected most of the dofs. In addition, the optimal $\alpha$ values, which correspond to the horizontal coordinates of the large dots, first decrease in value as $n$ goes from 1 to 5, then increase in value as $n$ increases to 20. This effect indicates that a tradeoff exists between the use of the relay and the amount of wasted retransmissions by the source.

Although not explicitly shown here, we can plot and compare the expected completion time per data packet when the channel between $s$ and $d$ varies. It can be observed that when $p_{sd}$ increases, $s$ is used a larger fraction of the time, with $\alpha$ becoming 1 if $p_{sd}$ is larger than $p_{rd}$, similar to the coding at both nodes case discussed in Section~\ref{subsec:rs}.

From the above numerical evaluations, we can conclude that RLNC at the relay $r$ only while operating in a rateless fashion is not an efficient transmission scheme in terms of throughput. Figure~\ref{fig:rOnly_n} shows that using ARQ without coding ($n=1$, $\alpha=1$) achieves the best expected completion time, or the best throughput. However, one issue with the $n=1$ case is that each data packet, when transmitted successfully, requires an acknowledgement from $d$, i.e., $E/n=TE_{use}+E_{ack}$. Such frequent feedbacks are not energy efficient. If $n>1$, even though the effect of the $E_{ack}$ term is mitigated by amortization over a larger $n$, the large increase in the value of $T/n$ shown in Figure~\ref{fig:rOnly_n} indicates that coding at the relay only is the most energy efficient when $n=2$. In Section~\ref{subsec:comparison}, we shall compare the energy use of this particular case with other schemes.

\subsection{RLNC at the Source $s$ Only}

\begin{figure}[t!]
\centering
  \includegraphics[width=3.2in]{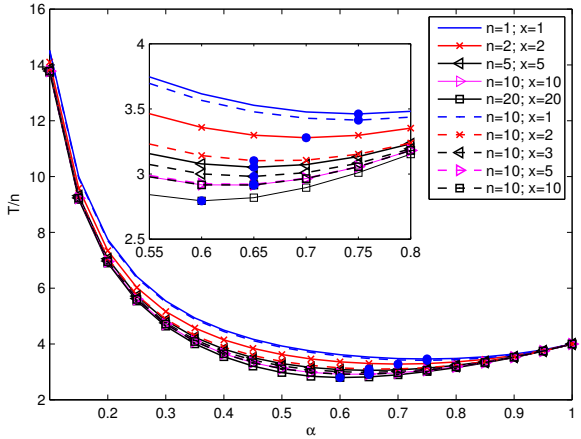}
  \caption{Coding at the source $s$ only, expected completion time per packet $T/n$  vs. $\alpha$, as $n$ and $x$ changes in value; $p_{sd} = 0.25$, $p_{sr}=0.8$, $p_{rd} =0.8$. The optimal $T^*/n$ is labeled with a large dot on each curve.}
  \label{fig:sOnly_n}
\end{figure}

Figure \ref{fig:sOnly_n} plots the expected completion time per data packet as a function of $\alpha$, when $n$ and $x$ vary. $n$ is the number of data packets to be transmitted by $s$, and $x$ is the amount of memory available at $r$ to store received mixtures. Unlike the coding at $r$ case, here $T/n$ decreases as $n$ becomes larger, because each packet sent by $s$ is innovative relative to $r$ and $d$, and as more packets are combined, the probability that a mixture sent by $r$ is innovative becomes larger. In addition to reducing $T/n$, another advantage of coding $n$ packets together at $s$ is that the cost for feedback can be amortized over a large number of data packets. Also observe from this figure that as low as $x=3$ units of memory suffices to achieve the expected completion time of the full memory case (i.e., $x=n$).

\begin{figure}[t!]
\centering
  \includegraphics[width=3.2in]{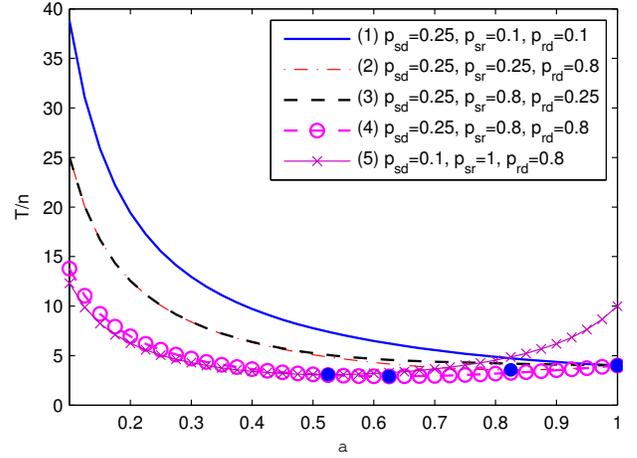}
  \caption{Coding at the source $s$ only, expected completion time per packet $T/n$ vs. $\alpha$, as $p_{sr}$ and $p_{rd}$ change; $p_{sd} = 0.5$, $n=10$, $x=n$.}
  \label{fig:sOnly_psrprd}
\end{figure}

Figure~\ref{fig:sOnly_psrprd} plots the expected completion time per data packet as a function of $\alpha$, for different $p_{sd}$, $p_{sr}$, and $p_{rd}$. Comparison among curves (1), (4) and (5) show that $r$ should be given more time to transmit when the tandem link from $s$ to $d$ through $r$ is more reliable than the direct link between $s$ and $d$. Comparison between (3) and (4), however, show that $r$ should not be used if the channel between $r$ and $d$ sees large packet losses, even if the channel between $s$ and $r$ is relatively reliable. This observation echoes the decision of not using the relay in the full coding case, as given by Eq.~\eqref{eq:alpha}, and discussed in Section~\ref{subsec:rs}. Moreover, comparison among curves (2), (3), and (4) show that the optimal value of $\alpha$ is a function of channel conditions.

\subsection{Comparisons}\label{subsec:comparison}

\begin{figure}[t!]
  \centering
    \includegraphics[width=3.3in]{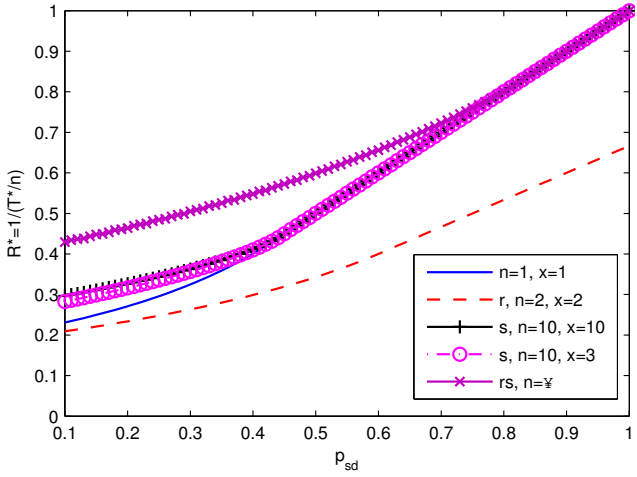}
    \caption{Achievable throughput as a function of $p_{sd}$, $R^*=\frac{1}{T^*/n}$; $p_{sr}=0.8$, $p_{rd}=0.8$.}
    \label{fig:comparison_psd}
\end{figure}

\begin{figure}[t!]
\centering
  \includegraphics[width=3.3in]{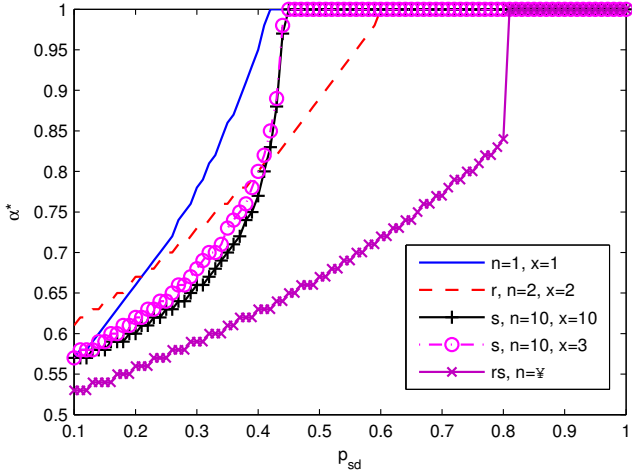}
  \caption{Optimal $\alpha^*$ corresponding to throughput values in Figure~\ref{fig:comparison_psd}; $p_{sr}=0.8$, $p_{rd}=0.8$.}
  \label{fig:comparison_psd_alpha}
\end{figure}

Figure~\ref{fig:comparison_psd} compares the maximum achievable rates of three cases: coding at $r$ only as discussed in Section~\ref{subsec:r}, coding at $s$ only as discussed in Section~\ref{subsec:s}, and coding at both $s$ and $r$ as discussed in Section~\ref{subsec:rs}. Figure~\ref{fig:comparison_psd_alpha} plots the corresponding $\alpha^*$ values that achieve these rates. For the coding at $r$ and coding at $s$ cases, the metric being plotted is the inverse of the optimal expected transmission completion time per data packet ($T^*/n$). This inverse corresponds to the throughput $R^*$ of the systems under discussion. For the case where coding is performed at both $s$ and $r$, the achievable rate is computed using Equations (1) and (2).

When RLNC is performed at $r$ only, as previous discussions have suggested, it is more desirable to mix fewer number of packets; since packets retransmitted from $s$ are uncoded, a larger fraction of the repetitions are wasted. In Figure~\ref{fig:comparison_psd}, the achievable rates are given for two different values of $n$. When $n=1$, coding is not performed, hence the transmission degenerates into a routing scheme: $s$ and $r$ retransmit a single packet until an acknowledgement is received from $d$. Observe from Figure~\ref{fig:comparison_psd_alpha} that when the channel between $s$ and $d$ is poor (e.g. $p_{sd}=0.2$), the route through $r$ is preferred ($\alpha^* \sim 0.65$), otherwise $r$ is not used ($\alpha^* = 1$). When $n=2$, $r$ still performs network coding, but only as the sum of two packets. Recall the assumption that all mixed packets transmitted by $r$ are innovative relative to $d$; the second curve (`$r, n=2, x=2$') in Figure~\ref{fig:comparison_psd} is therefore an upper bound on the actual system throughput, reconfirming that coding at $r$ only is not throughput efficient.

When RLNC is performed at $s$ only, Figure~\ref{fig:comparison_psd} shows that more than $69\%$ of the rate attained by the coding at both nodes scheme can be achieved. Here the achievable rates are plotted for only one set of channel realizations, with $p_{sr}=0.8$ and $p_{rd}=0.8$. The exact amount of coding gain depends on the reliability of all three links in the relay channel. Also observe that the performance gap decreases as the channel between $s$ and $d$ becomes more reliable. Moreover, Figure~\ref{fig:comparison_psd_alpha} shows that, when coding at $s$ only, transmissions from $r$ are not required after $p_{sd}$ becomes reasonably good (e.g., $p_{sd} > 0.442$). This is because transmissions from $r$ follow a randomized scheme, leading to redundant repetitions that do not contribute additional dof to $d$.

\begin{figure}[t!]
  \centering
    \includegraphics[width=3.3in]{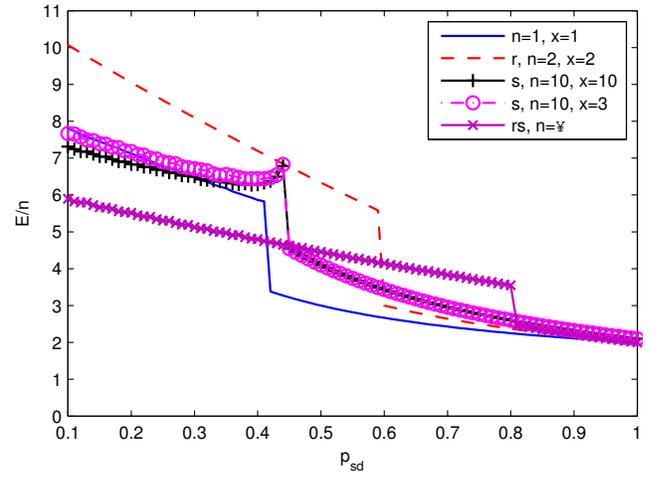}
    \caption{Packet delivery energy $E/n$ as a function of $p_{sd}$, corresponding to the optimal $\alpha^*$ in Figure~\ref{fig:comparison_psd}; $p_{sr}=0.8$, $p_{rd}=0.8$, $E_{tx}=1$, $E_{rx}=1$, $E_{nc}=1$, $E_{ack}=1$.}
    \label{fig:comparison_ET_ack_1_1_1_1_E}
 \end{figure}
\begin{figure}[t!]
\centering
  \includegraphics[width=3.3in]{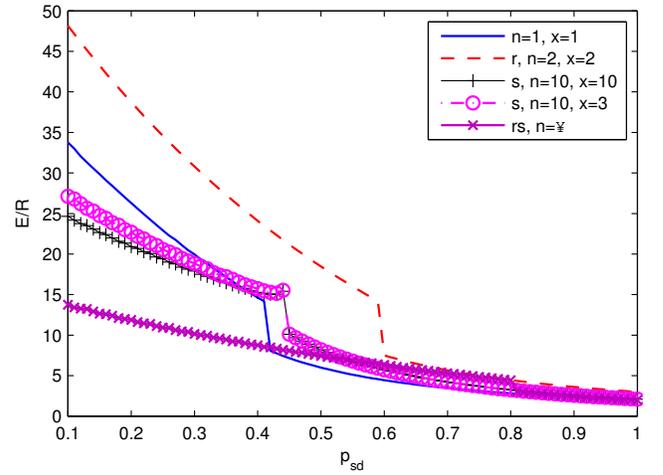}
  \caption{Packet delivery energy per throughput rate $E/(nR)$ as a function of $p_{sd}$, corresponding to the optimal $\alpha^*$ in Figure~\ref{fig:comparison_psd}; $p_{sr}=0.8$, $p_{rd}=0.8$, $E_{tx}=1$, $E_{rx}=1$, $E_{nc}=1$, $E_{ack}=1$.}
  \label{fig:comparison_ET_ack_1_1_1_1_ER}
\end{figure}

In maximizing throughput, coding as much as possible while fully using the relay $r$ seems to be the optimal strategy, followed by coding at $s$ alone. However, assuming that both coding and listening costs power, such approaches may pay higher costs in terms of energy. As discussed in Section~\ref{sec:analysis}, our analysis enables the derivation of total energy costs. For example, if $E_{tx}$, $E_{rx}$, $E_{nc}$, $E_{ack}$ are identically 1, Figure~\ref{fig:comparison_ET_ack_1_1_1_1_E} plots the packet deliver energy corresponding to the optimal $\alpha^*$ in Figure~\ref{fig:comparison_psd_alpha}, while Figure~\ref{fig:comparison_ET_ack_1_1_1_1_ER} plots this energy consumption scaled by the maximum achievable rate. The different energy terms have been chosen assuming that coding and listening consumes energy on the same scale as transmission. Such assumptions are valid in systems where just having the circuitry turned on constitutes the most significant portion of energy use. Other ranges of values are also possible, as we have discussed in \cite{shi2011both}, depending on the underlying physical layer hardware implementations.

It is easy to see from these figures that when $p_{sd}$ is low, coding at both $s$ and $r$ is the most throughput and energy efficient, while coding at $s$ alone provides a compromise between throughput and energy use; under better channel conditions, however, not coding ($n=1$) and not using the relay ($\alpha=1$) require energy, while achieving equally good throughputs. At $p_{sd}=1$, the energy cost for the successful delivery of one data packet is 2 when coding is conducted at both $s$ and $r$: one on transmission, and one on coding. On the other hand, the energy cost for coding at $s$ only, assuming $n=10$, is $2.1$: one for transmission, one for coding, and $1/10$ for listening to transmission termination acknowledgement. Moreover, the energy cost for coding at $r$ only, assuming $n=2$, is $2$: according the coupon collector's problem, on average 3 units of energy are spent on transmitting the 2 packets, and one unit of energy is spent on receiving the acknowledgement. Lastly with simple ARQ ($n=1$), two units of energy are spent on each successfully delivered packet.
\begin{figure}[t!]
\centering
  \includegraphics[width=3.3in]{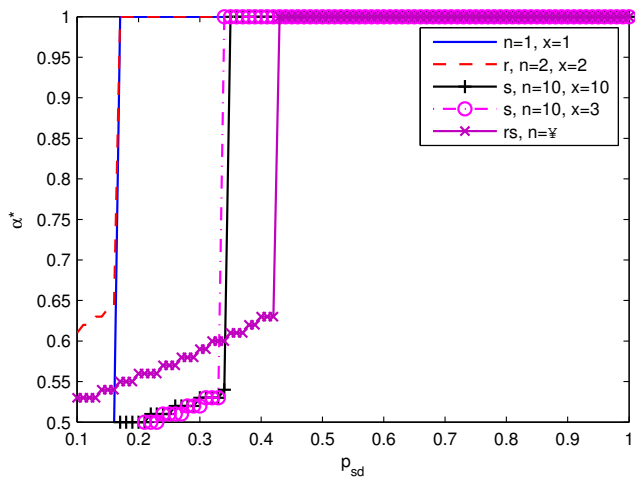}
  \caption{Optimal $\alpha^*$ corresponding to packet delivery energy values in Figure~\ref{fig:comparison_EE_ack_1_1_1_1_E}; $p_{sr}=0.8$, $p_{rd}=0.8$, $E_{tx}=1$, $E_{rx}=1$, $E_{nc}=1$, $E_{ack}=1$.}
  \label{fig:comparison_EE_ack_1_1_1_1_alpha}
\end{figure}
\begin{figure}[t!]
  \centering
    \includegraphics[width=3.3in]{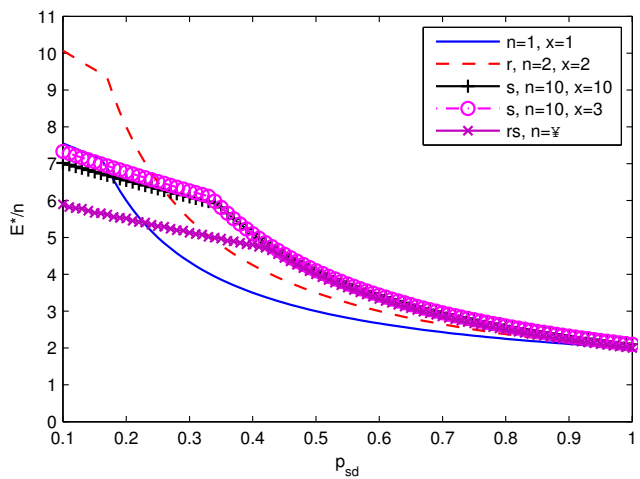}
    \caption{Minimum packet packet delivery energy $E^*/n$ as a function of $p_{sd}$, $p_{sr}=0.8$, $p_{rd}=0.8$, $E_{tx}=1$, $E_{rx}=1$, $E_{nc}=1$, $E_{ack}=1$.}
    \label{fig:comparison_EE_ack_1_1_1_1_E}
\end{figure}
Under the same channel conditions and system parameters as given in Figure~\ref{fig:comparison_ET_ack_1_1_1_1_E}, optimizing for energy use leads to a very different set of $\alpha$ values, plotted in Figure~\ref{fig:comparison_EE_ack_1_1_1_1_alpha}. The corresponding optimal packet delivery energies are shown in Figure~\ref{fig:comparison_EE_ack_1_1_1_1_E}. Observe that the decision to turn off the relay $r$ entirely comes at smaller $p_{sd}$ values. This is because $r$ consumes energy in listening to incoming packets from $s$ as well as sending outgoing packets to $d$. The energy cost of using $r$ is the same as retransmitting twice from $s$. In addition, since $r$ shares the use of the wireless medium with $s$, having $r$ turned on reduces the rate at which packets can be transmitted from $s$. With these two effects combined, $r$ is used only at small $p_{sd}$ values. Another result of the energy tradeoff between $s$ and $r$ observable from these two figures is that even though the optimal packet deliver energy curve is continuous, $\alpha^*$ sees a jump for each of the coding strategies.

Similar energy and throughput curves can be evaluated when energy parameters $E_{tx}$, $E_{rx}$, $E_{nc}$ and $E_{ack}$ take on different ranges. In practical systems, depending on the underlying circuit implementation, one or more of these energy terms can dominate over the others, and the optimal transmission schedule could be very different from the ones shown above. Nonetheless, our analysis enables robust decision making to determine when and where to code in a wireless packet erasure relay channel.

\section{Conclusion}\label{sec:conclusion}
We propose Markov chain models to analyze the throughput and packet delivery energy performances of network coding strategies in the wireless packet erasure relay channel. The evolution of innovative packets are tracked when either or both the source and the relay perform random linear network coding. We show through numerical evaluations that using a random code at the relay alone is neither throughout nor energy efficient, while coding at the source alone can provide a good tradeoff between throughput and energy use. We also show that only a very small amount of memory is required at the relay when coding is performed at the source only. Although we do not attempt to categorize explicitly the optimal network coding strategies in the relay channel under different system parameters, we provide a framework for deciding whether and where to code, taking into account of throughput maximization and energy depletion. Future work will consider the use of systematic codes, which have been mentioned in this paper but not studied in detail. A natural extension of the three-node relay channel is a star-shaped network, where nodes can  act as relays for their neighbors. A direct generalization of our given framework does not seem tractable, nonetheless it is clear from our short analysis that the problem of choosing an optimal coding subgraph is very important when practical constraints, such as energy, are taken into account.


\bibliographystyle{IEEEtran}
\bibliography{references}

\begin{IEEEbiography}[{\includegraphics[width=1in,height=1.25in,clip,keepaspectratio]{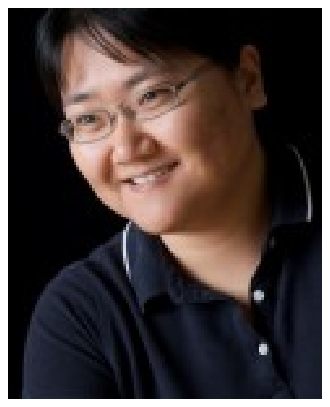}}]
{Xiaomeng Shi}
is currently pursuing a Ph.D. degree in Electrical Engineering and Computer Science at the Massachusetts Institute of Technology (MIT), Cambridge, USA. She received a B.Eng. degree in Electrical and Computer Engineering in 2005 from the University of Victoria, Victoria, BC, Canada, and an S.M. degree in Electrical Engineering in 2008 from MIT. Her research interests include network coding, energy efficient protocol design in wireless networks, and signal processing.
\end{IEEEbiography}

\begin{IEEEbiography}[{\includegraphics[width=1in,height=1.25in,clip,keepaspectratio]{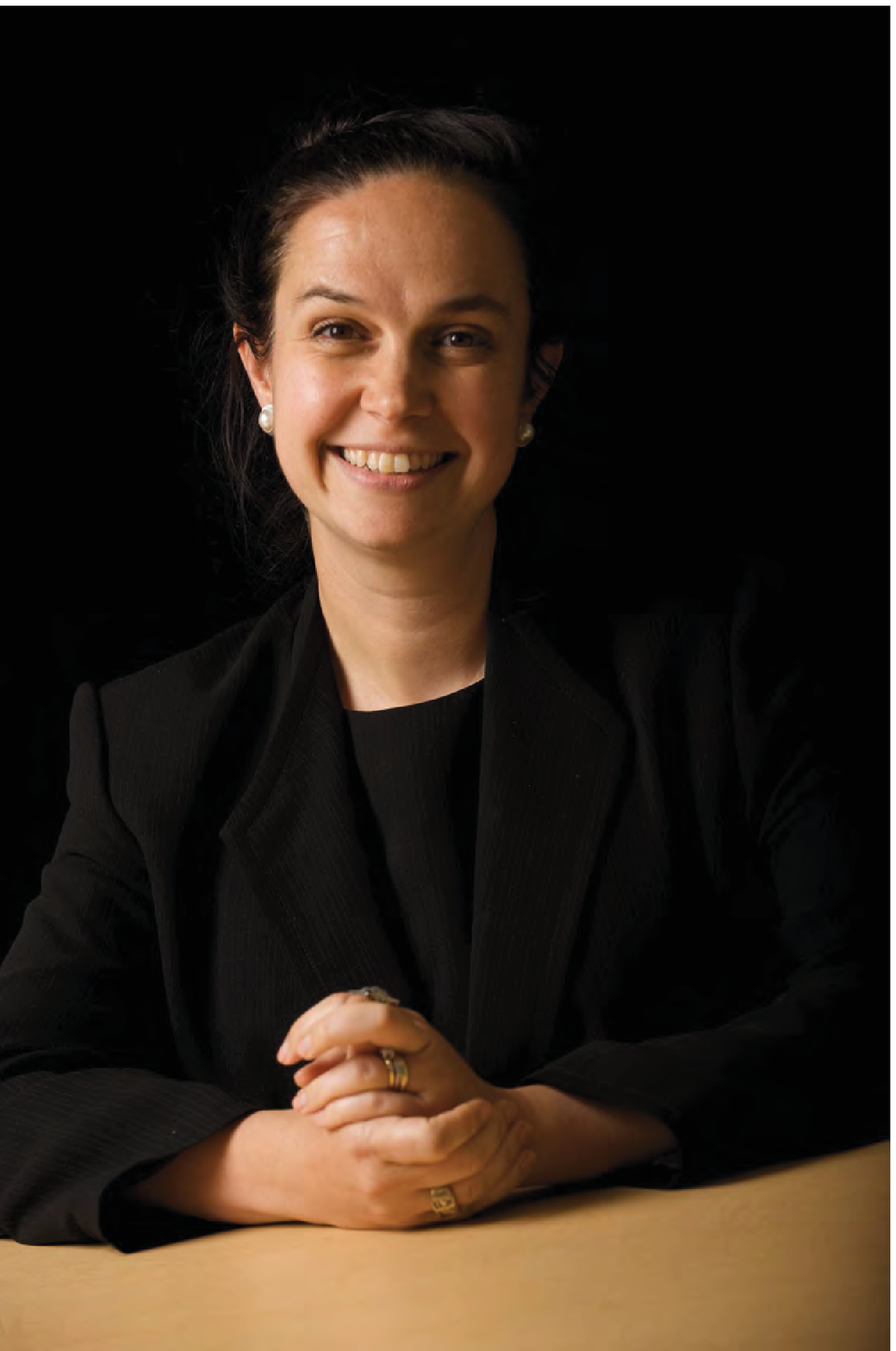}}]
{Muriel M\'edard} is a Professor of Electrical Engineering at MIT. She was previously an Assistant Professor in the ECE Department at UIUC and a Staff Member at MIT Lincoln Laboratory. She received B.S. degrees in EECS, in Mathematics, and in Humanities, as well as M.S. and Sc D. degrees in EE, all from MIT. She has served as an Associate Editor for the Optical Communications and Networking Series of the IEEE Journal on Selected Areas in Communications, the IEEE Transactions on Information Theory and the OSA Journal of Optical Networking. She has served as a Guest Editor for the IEEE Journal of Lightwave Technology, the IEEE Transactions on Information Theory
(twice), the IEEE Journal on Selected Areas in Communications and the IEEE Transactions on Information Forensic and Security. She serves as an associate editor for the IEEE/OSA Journal of Lightwave Technology. She is a member of the Board of Governors of the IEEE Information Theory Society and serves as the President. She has served as TPC co-chair of ISIT, WiOpt and CONEXT. She was awarded the 2009 IEEE Communication Society and Information Theory Society Joint Paper Award , the 2009 IEEE William R. Bennett Prize in the Field of Communications, and the 2002 IEEE Leon K. Kirchmayer Prize Paper Award. She was co-winner of the 2004 MIT Harold E. Edgerton Faculty Achievement Award. In 2007, she was named a Gilbreth Lecturer by the National Academy of Engineering. Professor M\'edard's research interests are in the areas of network coding and reliable communications, particularly for optical and wireless networks.
\end{IEEEbiography}

\begin{IEEEbiography}[{\includegraphics[width=1in,height=1.25in,clip,keepaspectratio]{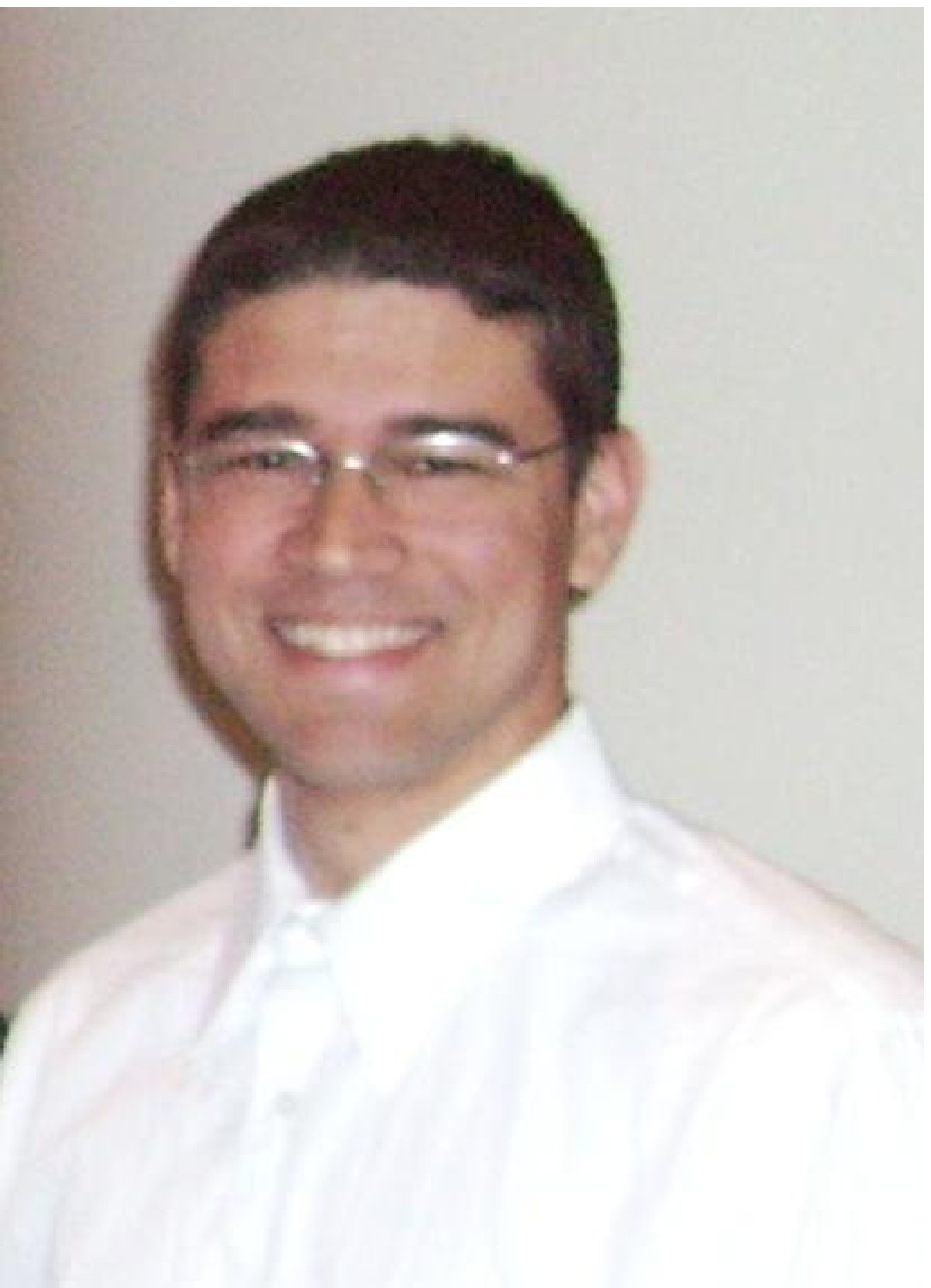}}]
{Daniel E. Lucani} is an Assistant Professor at the Faculty of Engineering of the University of Porto and a member of the Instituto de Telecomunica\c c\~oes (IT). He received his B.S.  (\textit{summa cum laude}) and M.S. (with honors) degrees in Electronics Engineering from Universidad Sim\'on Bol\'ivar, Venezuela in 2005 and 2006, respectively, and the Ph.D. degree in Electrical Engineering from the Massachusetts Institute of Technology (MIT) in 2010. His research interests lie in the general areas of communications and networks, network coding, information theory and their applications to highly volatile wireless sensor networks, satellite and underwater networks, focusing on issues of robustness, reliability, delay, energy, and resource allocation. Prof. Lucani was a visiting professor at MIT. He is the general co-chair of the Network Coding Applications and Protocols Workshop (NC-Pro 2011) and has also served as reviewer for high impact international journals and conferences, such as, the IEEE Journal of Selected Areas in Communications, IEEE Transactions on Information Theory, IEEE Transactions on Communications, IEEE Transactions on Wireless Communications, IEEE Journal of Oceanic Engineering.
\end{IEEEbiography}

\end{document}